\documentclass[aps,twocolumn]{revtex4}
\usepackage{epsfig}

\begin{document}

\title{Creation-annihilation  processes in the ensemble of constant
  particle number}

\author{Carlos E. Fiore and M\'{a}rio J. de Oliveira}

\affiliation{Instituto de F\'{\i}sica\\
Universidade de S\~{a}o Paulo\\
Caixa Postal 66318\\
05315-970 S\~{a}o Paulo, S\~{a}o Paulo, Brazil}
\date{\today}

\begin{abstract}

We study, in  the ensemble of constant 
particle number,  processes in which a cluster of particles is
annihilated and particles are created catalytically in active sites. 
In this ensemble, particles 
belonging to a  cluster of $\ell$ particles  jump to $\ell$ 
distinct active sites.
As examples of our prescription, we analyze numerically 
three nonequilibrium systems  that annihilate cluster of particles
that are  identified as conserved versions of the pair annihilation
contact model, triplet annihilation contact model and  pair 
contact process. We show also how to set up the constant particle
number ensemble from the constant rate ensemble.  

PACS numbers: 05.70.Ln, 05.50.+q, 05.65.+b

\end{abstract}

\maketitle

\section{Introduction}
The use of distinct ensembles in equilibrium statistical mechanics,
in which  systems are described by a given  Hamiltonian,
is a  well established  concept \cite{landau}. There is a standard procedure
for passing from a given ensemble to another.
For nonequilibrium systems \cite{marro},
on the other hand, this procedure can not be used, 
since they are not described by a Hamiltonian and therefore, their 
probability distribution are not known a priori.
In  most  cases nonequilibrium systems are defined in 
a constant rate  ensemble, (which we call ordinary version),
but the possibility of 
using another ensemble was put forward by Ziff and Brosillow
\cite{brziff} when they  used a constant coverage ensemble 
in their study of an irreversible surface-reaction model. 
Subsequently, Tom\'e and de Oliveira \cite{tome2001}
introduced the contact process in the ensemble of constant number
of particles.

In the conserved contact process (CCP)  \cite{tome2001}, a particle  chosen at
random leaves its place and jumps to one of the many
active sites of the lattice. The CCP
displays properties that in the thermodynamic
limit are identical to the ordinary contact process.
Hilhorst and van Wijland \cite{hilhorst} provided a proof
of the equivalence between the constant rate and the 
constant particle number ensembles for the contact process. Later,
de Oliveira \cite{mario2003} extended the proof
for any reaction process that annihilate one particle
 \cite{tome2001,sabag,fiore}. 
Sometimes the use of the conserved version
is more appropriate than the ordinary 
version as for example in the study of a first-order transition. 
This advantage has been exploited by Ziff and Brosillow \cite{brziff}, 
Loscar and Albano \cite{loscar} that studied hysteretic effects
in a model that describes the CO+NO  reaction,
and  more recently by Fiore and de Oliveira \cite{fiore}.

Here, we analyze in the ensemble of constant particle number 
one-dimensional 
models whose ordinary versions have been  previously studied. These
models are: the pair-annihilation contact model (PAM) \cite{dic89a},
the triplet-annihilation contact model (TAM) \cite{dic89a,dick1990},
and the pair contact process (PCP) \cite{jensen1993,jensend1993,dic02}.
In these  models, a cluster of particles is spontaneously annihilated
and  particles are catalytically created in  active sites. 
Active sites are empty sites surrounded by a neighborhood of
particles.

\section{TRANSITION RATES}

Consider a site $i$ of regular lattice. To each site $i$ 
of the lattice we attach an occupation variable $\eta_i$
which takes the values $0$ or $1$ according whether
the site $i$ is empty or occupied by a particle.
In the constant rate ensemble, 
the usual process is composed of creation
of  a single particle (0 $\rightarrow$ 1)  with transition rate 
$w_i^c=k_{c}\omega_{i}^{c}$, 
and annihilation of a cluster of $\ell$ particles
in a row
(111...1$\rightarrow$000...0) with transition rate
$w_i^a=k_{a}\omega_{i}^{a}$.
The transition rate $w_i^c$ is the probability per unit time 
of creating a particle at the site $i$.
The transition rate $w_i^a$ is the probability per unit time 
of annihilating a particle at the site $i$.
The total transition rate for these two reaction processes
 is given by
\begin{equation}
\label{rateord}
 w_{i} = w_i^a +w_i^c
= k_{a}\omega_{i}^{a}+k_{c}\omega_{i}^{c}.
\end{equation}
The quantities $\omega_{i}^{a}$ and $\omega_{i}^{c}$ will be defined
according to the specific model and they depending on the local
configuration of particles. The quantities $k_{a}$ and $k_{c}$, which
we call strengths of the annihilation and creation process,
are parameters  that gives the weight of each subprocess.
Diffusion of particles consists of a particle hopping to its nearest
neighbor. 
In diffusive models, the diffusion of particles and 
the reaction process occur with probability
$D'$ and $1-D'$, respectively.

In the ensemble of constant particle number, the jumping process
and the diffusion of particles are chosen with probabilities 
$1-D$ and $D$, respectively. In the hopping step, an
occupied site $\eta_{i}=1$ and its nearest next neighbor
$\eta_{j}=0$ interchange their occupation variables, whereas
in the jumping step,  
$\ell$ adjacent particles leave their places and arrive at
$\ell$ active sites. 
Thus, in the ensemble of constant particle number, both the creation
and annihilation of particles are replaced by $\ell$ jumping process.

\section{EQUIVALENCE OF ENSEMBLES}

In the construction of conserved models 
we have to be  concerned only with the creation and annihilation processes
since the diffusion process already conserves the particle number.
In the case of
processes that annihilate a pair of particles (PAM and PCP), the
jumping process occurs with rate  $\omega_{i}^{a}\omega_{j}^{c}\omega_{k}^{c}$.
In the case of  annihilation of a triplet of neighboring particles (TAM), it
occurs with rate  $\omega_{i}^{a}\omega_{j}^{c}\omega_{k}^{c}\omega_{m}^{c}$.

To demonstrate that this conserved dynamics is equivalent to the
ordinary dynamics described
by Eq. (\ref{rateord}) we follow the same reasoning given by
de Oliveira \cite{mario2003}.
Let us consider the specific case of models that annihilate 
two particles.
In this case, two  particles, belonging to the neighborhood of  site $i$, 
chosen at random, jump to two
distinct active sites, $j$ and $k$, also chosen at random. 
The jumping process  occurs with rate
$\omega_{i,j,k}=\omega_{i}^{a}\omega_{j}^{c}\omega_{k}^{c}/L^{2}$
where $L$ 
is the number of sites
of the lattice. Let us evaluate the  total rate $\sum_{j,k}\omega_{i,j,k}$
in which particles leave the neighborhood of site $i$. In the
thermodynamic limit, the sums
$\sum_{j}\omega_{j}^{c}/L$ and $\sum_{k}\omega_{k}^{c}/L$ approach,
by the law of large numbers, the averages 
$\langle \omega_{j}^{c} \rangle$ and $\langle \omega_{k}^{c} \rangle$,
respectively.
Therefore we have 
$\sum_{j,k}\omega_{i,j,k}=\langle \omega_{j}^{c} \rangle \langle \omega_{k}^{c} 
\rangle \omega_{i}^{a}$.
Similarly, the total rate at which a particle arrives at the site $j$
is $\langle \omega_{i}^{a} \rangle \langle \omega_{k}^{c} 
\rangle \omega_{j}^{c}$
and  the total rate at which a particle gets 
 to the site $k$ is $\langle \omega_{i}^{a} \rangle 
\langle \omega_{j}^{c} 
\rangle \omega_{k}^{c}$.

Comparing these results with the rate (\ref{rateord}), 
we see that $k_c$ and $k_a$ should be proportional to
$2 \langle \omega_{i}^{a} \rangle \langle \omega_{k}^{c} \rangle $ 
and $\langle \omega_{j}^{c} \rangle \langle \omega_{k}^{c} \rangle $, respectively.
Defining $\alpha$ as the ratio $k_a/k_c$, we can write the following relation 
\begin{equation}
\label{equiv}
\alpha=\frac{\langle \omega_{j}^{c}\rangle}
{2\langle \omega_{i}^{a}\rangle},
\end{equation}              
valid for any  process 
that annihilates a pair of particles in the ensemble
of constant particle number.

When we add a diffusive step, the ensembles  are also 
equivalent, but 
the  diffusion rate $D'$ used in the constant rate ensemble
will not have, in general, 
 the same value of diffusion rate $D$ that is used
in the constant particle number ensemble. A  relation
between the  rates $D'$ and  $D$ is given by
\begin{equation}
\label{equivd}
\frac{1-D'}{D'}=2\frac{1-D}{D}\frac {\langle \omega_{j}^{c}\rangle}{\rho}
 \frac {\langle \omega_{i}^{a}\rangle}{\rho}.
\end{equation}
where the factor $1/\rho^{2}$ comes from the 
ratio between the occurrences of 
diffusive process and the jumping process. 

Generalizing  for  an
arbitrary cluster of  $\ell$ particles, the Eqs. (\ref{equiv})
 and (\ref{equivd}) become
\begin{equation}
\label{equivgeral}
\alpha=\frac{\langle \omega_{j}^{c}\rangle}
{\ell\langle \omega_{i}^{a}\rangle},
\end{equation}
and
\begin{equation}
\label{equi}
\frac{1-D'}{D'}=\frac{1-D}{D}\ell\langle \omega_{j}^{c}
\rangle^{\ell-1}\langle \omega_{i}^{a}\rangle\rho^{-\ell},
\end{equation}
respectively. In the
particular case of $\ell=1$, we have $D=D^{'}$ because
$\langle \omega_{i}^{a}\rangle=\langle \eta_{i}\rangle= \rho$,
as it has already been obtained and used
in the simulations of diffusive contact processes
\cite{mario2003,fiore}.
 
\section{Pair-annihilation contact model}

\subsection{Constant rate ensemble}

In the ordinary PAM, the creation of particles
is catalytic and a pair of
particles is annihilated spontaneously. It is represented by the 
chemical reactions:
\begin{equation}
0+A \rightarrow A+A,
\end{equation}
\begin{equation}
A+A \rightarrow 0+0,
\end{equation}
describing the catalytic 
creation and annihilation of a pair of particles, respectively.
The rates $\omega_{i}^{c}$ and
$\omega_{i}^{a}$ in Eq. (\ref{rateord}) are given by
\begin{equation}
\label{rate1}
\omega_{i}^{c}=(1-\eta_{i})\frac{1}{z}\sum_{\delta}\eta_{i+\delta},
\end{equation}
where the summation is over the $z$ nearest neighbor sites
and
\begin{equation}
\label{rate2}
\omega_{i}^{a}=\eta_{i}\eta_{i+1}.
\end{equation}
The strengths of the creation and annihilation
processes are given by $k_{c}=1$ and $k_{a}=\alpha$.

For values of $\alpha > \alpha_{c}$, the system is constrained into the
absorbing state (without particles), whereas for
$\alpha < \alpha_{c}$ we have an active state in which particles
are created and pairs of neighboring particles are annihilated. A continuous
phase transition between these two regimes occurs at $\alpha=\alpha_{c}
=0.18622(3)$ \cite{dic89a}. Close to the critical point, the 
order parameter (here the density of particles $\rho$) follows
 a power law 
\begin{equation}
\label{powerlaw}
\rho  \sim (\alpha_{c}-\alpha)^{\beta},
\end{equation}
where $\beta=0.2765(1)$. The PAM is found to 
belong to  the directed percolation universality class (DP)
\cite{marro}.

\subsection{Constant particle number ensemble}

\begin{figure}
\setlength{\unitlength}{1.0cm}
\includegraphics[scale=0.5]{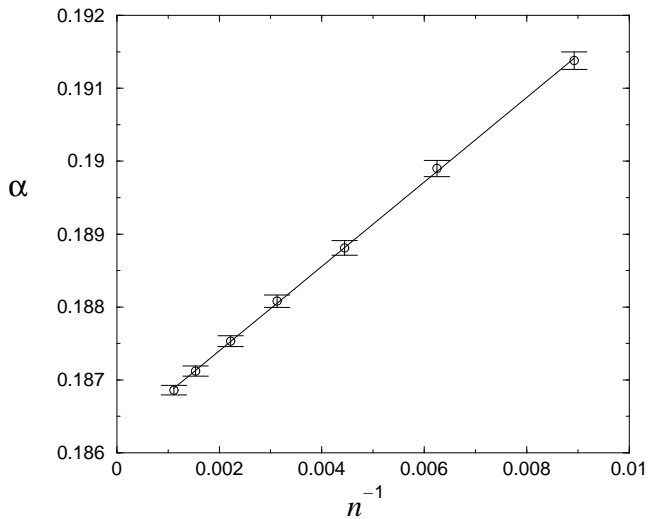}
\caption{Values of $\alpha$ versus the number of particles $n$  for
  the conserved PAM in the absence of diffusion
for an infinite system. The line corresponds to the linear 
extrapolation using Eq. (\ref{expsub}).}
\label{CPAM}
\end{figure}

The Monte Carlo simulation of the conserved PAM
is performed as follows. The jumping process and the diffusive process
are chosen with probabilities $1-D$ and $D$, respectively. If the 
jumping process is chosen, then
 a pair of neighboring particles
is chosen at random. Next, we choose  two empty sites each one surrounded by at
least one particle. The two particles  belonging to the selected pair
jump to these two active sites. If the diffusive process is chosen, 
then a particle chosen randomly is moved to one of its neighboring
site, provided it is empty.
In this ensemble, the rate
$\alpha$ is evaluated by using  formula (\ref{equiv}), where
the quantities $ \omega_{i}^{a} $ and  $ \omega_{i}^{c} $
are given by Eqs. (\ref{rate1}) and (\ref{rate2}).
To determine the critical rate $\alpha_c $ we have simulated the conserved PAM
 in the 
subcritical regime. In this  regime we 
have an infinite lattice and a finite particle number $n$.
The critical value $\alpha_c$ is obtained by assuming 
the asymptotic relation
\begin{equation}
\label{expsub}
\alpha -\alpha_c \sim \frac{1}{n}.
\end{equation}
A linear extrapolation of $\alpha$ versus $ n^{-1}$ gives  $\alpha_c$,
since when $n \rightarrow \infty$ we have $\alpha \rightarrow \alpha_c$.
The behavior of $\alpha$ versus
$n^{-1}$  for the conserved PAM in the subcritical regime is shown
in  Fig. \ref{CPAM}.  In the limit of
$n \rightarrow \infty$, we obtain 
$\alpha_{c}=0.18624(7)$, in excellent agreement 
with the value $\alpha_{c}=0.18622(3)$ obtained from its ordinary version
\cite{dic89a}. 

According to Vicsek \cite{vicsek}, 
in the critical point, we have a formation of fractal clusters.
To calculate the fractal dimension for 
a fixed $D$, we have measured
the maximum distance $R$ between two particles of the system
as a function of the particle number $n$.

\begin{figure}
\setlength{\unitlength}{1.0cm}
\includegraphics[scale=0.5]{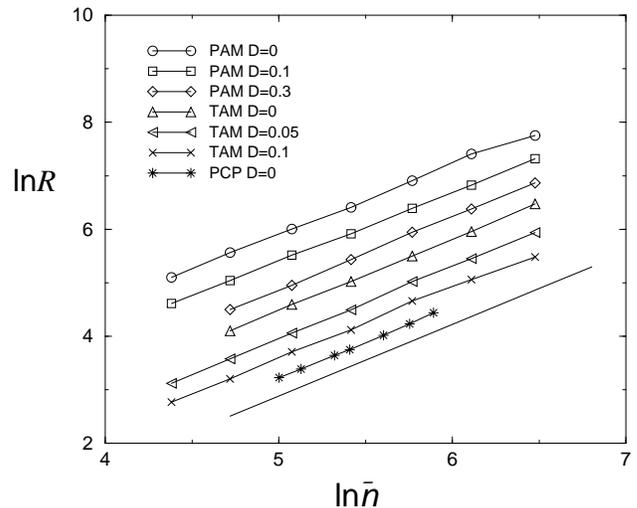}
\caption{Log-log plot of the size system $R$, in the subcritical regime,
as function of $\overline n$ where $\overline n$ is 
the particle number $n$ for the conserved PAM and
the conserved TAM, and the  number of pairs of neighboring particles 
$n_p$ for the conserved PCP. 
For comparison, we show a  straight line with slope $1.338$.}
\label{fractal}
\end{figure}

Following
Br\"oker and Grassberger \cite{brograss}, we assume
the following asymptotic behavior
\begin{equation}
R \sim  n^{1/d_{F}},
\end{equation}
where $d_{F}$ is the fractal dimension. The fractal dimension is related to 
the survival probability exponent $\delta$, the mean number of particles
exponent $\eta$, and the dynamic exponent $z$ by
$d_{F}=2(\delta+\eta)/z$ \cite{marro}.
In  Fig. \ref{fractal} we show a log-log plot of $R$ versus $n$
for the conserved PAM for some values of the diffusion rate $D$.
The values of $1/d_{F}$ are consistent with the value $1.338(6)$, 
obtained for the CCP \cite{sabag}.

\begin{figure}
\setlength{\unitlength}{1.0cm}
\includegraphics[scale=0.5]{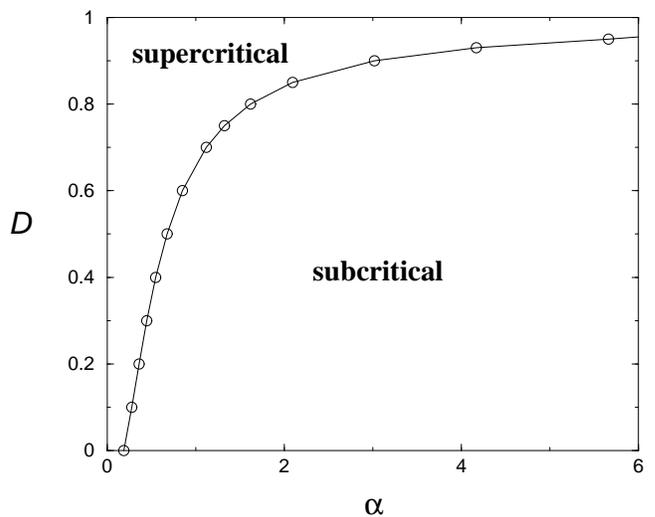}
\caption{Phase diagram for the conserved PAM in $D$ versus $\alpha$ space.
 The supercritical and subcritical regimes are separated by a critical line.}
\label{fig4}
\end{figure}

Using Eq. (\ref{expsub}) we have built the phase diagram 
shown in  Fig. \ref{fig4}
for several values of diffusion rate $D$.
Increasing the diffusion rate, we  
expect an increase in the value of $\alpha_c$, since 
by dispersing particles the density of pairs of neighboring particles
$\langle \omega_{i}^{a}\rangle$
decreases  whereas the density of active sites 
$\langle \omega_{i}^{c}\rangle$ increases.
Our results show that for sufficiently rapid diffusion, $D\rightarrow 1$,
the critical value of $\alpha_c$ increases without limit, that is,
$\alpha_c \rightarrow \infty$,
in agreement with the results obtained by Dickman \cite{dic89a}.

\begin{figure}
\setlength{\unitlength}{1.0cm}
\includegraphics[scale=0.5]{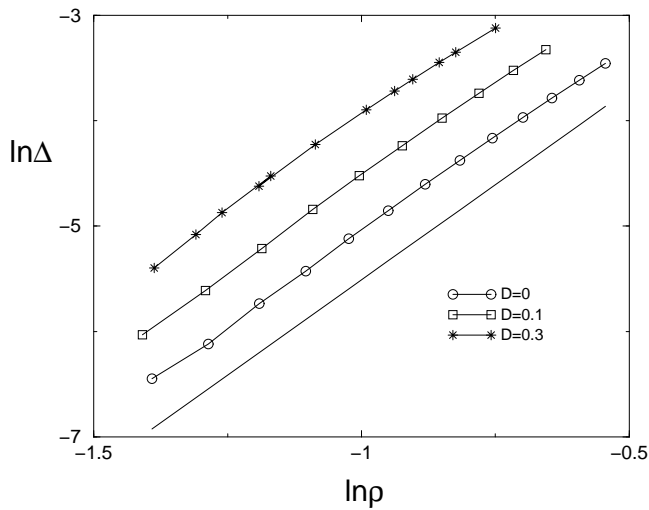}
\caption{log-log plot of $\Delta\equiv\alpha_{c}-\alpha$ versus
the density $\rho$, in the supercritical regime for the conserved PAM for
some values of diffusion rate. The straight line has a slope 3.61. }
\label{betapam}
\end{figure}

To determine the exponent $\beta$,   
we have simulated the conserved PAM in the 
supercritical regime for some values of the diffusion rate $D$.
In this regime,  the density $\rho=n/L$ is kept 
fixed for a size system $L$. We have used a lattice
with $L=10000$ sites and  a number of
Monte Carlo steps ranging from $10^{6}$ to $ 10^{7}$.
The exponent $\beta$ is obtained from the
log-log plot of $\Delta\equiv\alpha_c-\alpha$ versus $\rho$
as shown in the Fig. \ref{betapam}. 
We used the values of $\alpha_{c}$ calculated by
using the equation (\ref{expsub}).
For $D=0$, we obtained $1/\beta=3.61(3)$ 
that is in excellent agreement with 
the value  $1/\beta=3.61(1)$  obtained for the 
CCP \cite{tome2001}. The simulations performed at
$D=0.1$ and $D=0.3$ give exponents also
compatible with this value.

\section{Triplet-annihilation contact model}

\subsection{Constant rate ensemble}

The ordinary TAM is  composed of the creation of particles
 and annihilation of three neighboring 
particles. It is  represented by the chemical reactions:
\begin{equation}
0+A \rightarrow A+A,
\end{equation}
\begin{equation}
\label{scheme3}
A+A+A \rightarrow 0+0+0,
\end{equation}
describing the catalytic 
creation and annihilation of a triplet of particles, respectively.
 The quantity
 $\omega_{i}^{c}$ is also given by
Eq. (\ref{rate1}) and
the rate $\omega_{i}^{a}$ is given by
\begin{equation}
\label{rate3}
\omega_{i}^{a}=\eta_{i-1}\eta_{i}\eta_{i+1}.
\end{equation}
Without diffusion, a continuous
phase transition to an absorbing state
occurs at $\alpha_{c} =0.1488(2)$ \cite{dic89a,dick1990} with
critical exponent $\beta= 0.2765$.
 
The competition between diffusion and reaction  process may bring 
great changes in the phase diagram. For example, in the multiple-creation
contact processes, the transition becomes first order for 
high enough values of diffusion \cite{tome2001,fiore}. In the PCP, 
several works have reported a change in the universality 
class for nonzero values of diffusion rate.
For the ordinary TAM, previous works \cite{dic89a,dick1990}  have shown that 
in the presence of diffusion, the system
exhibits a reentrant phase diagram as can
be seen in  Fig. \ref{fig5}.

In references \cite{dic89a,dick1990},
the processes  of diffusion, creation and
annihilation are chosen  with probabilities $ D^{*}$,
$(1- D^{*})\lambda/(\lambda+1)$ and
$(1- D^{*})/(\lambda+1)$, respectively.
To compare  ordinary  results with ours, it is  necessary to convert 
the parameters. The relation between
$D^{*}$ and $D'$, $\lambda$ and $\alpha$ are given by
\begin{equation} 
\lambda=\frac{1}{\alpha},
\end{equation}
and
\begin{equation} 
\label{convert2}
 D^{*}=\frac{D'}{D'+(1+\alpha)(1-D')}.
\end{equation}

For values of the diffusion rate $D^{*}$
higher than  $D^{*}_{\rm max}=0.587$, there is
no phase transition and the system displays only active state.
For $D^{*}<D^{*}_{\rm max}$,  an active state is also possible for 
very small values of $\lambda$. This happens because there are 
few  isolated particles. Since the system annihilates only
triplets of neighboring particles, these sparse particles 
are able to ``survive''. 
On other hand, the probability that a new particle is created
is very low because $\lambda$ is very small.

\subsection{Constant particle number ensemble}
The Monte Carlo simulation of the conserved TAM is performed as
follows:
 The jumping process and the diffusive process
are chosen with probabilities $1-D$ and $D$, respectively. If the 
jumping process is chosen, then
 a triplet of neighboring particles
is chosen at random. Next, we choose  three
 empty sites each one surrounded by at
least one particle. The three particles  belonging to the selected triplet
jump to these three active sites. If the diffusive process is chosen, 
then a particle chosen randomly is moved to one of its neighboring
site, provided it is empty. 
The quantities $\omega_i^c$ and $\omega_{i}^{a}$ are calculated by
using the
Eqs. (\ref{rate1}) and (\ref{rate3}).
The rate $\alpha$ is evaluated  using  formula (\ref{equivgeral})
with $\ell =3$.

\begin{figure}
\setlength{\unitlength}{1.0cm}
\includegraphics[scale=0.5]{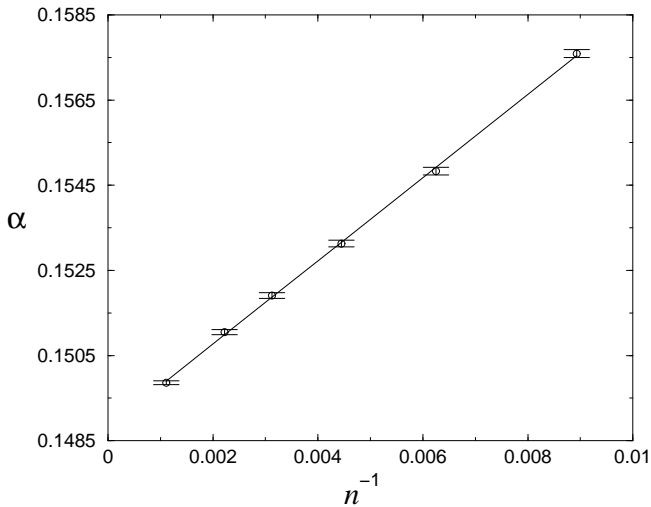}
\caption{Values of $\alpha$ versus the number of particles $n$  
for the conserved TAM
in the absence of diffusion
for an infinite system. The line corresponds to the linear 
extrapolation using Eq. (\ref{expsub}).}
\label{CTAM}
\end{figure}

To determine the critical value $\alpha_c$ we have simulated 
the conserved TAM in the subcritical regime. We assume a  behavior given 
by Eq. (\ref{expsub}) for large particle number $n$.
For $D=0$, the values of $\alpha$ versus  $n$ are shown 
in  Fig. \ref{CTAM}.
The linear extrapolation gives the critical rate
$\alpha_{c}=0.14898(5)$,  which agrees very well 
with the value $\alpha_{c}=0.1488(2)$ obtained for the ordinary version
\cite{dic89a}.

For the values of diffusion used here, we have also obtained 
values of  $1/d_{F}$ consistent with $1.338(6)$, as shown in the Fig.
\ref{fractal}.

\begin{figure}
\setlength{\unitlength}{1.0cm}
\includegraphics[scale=0.45]{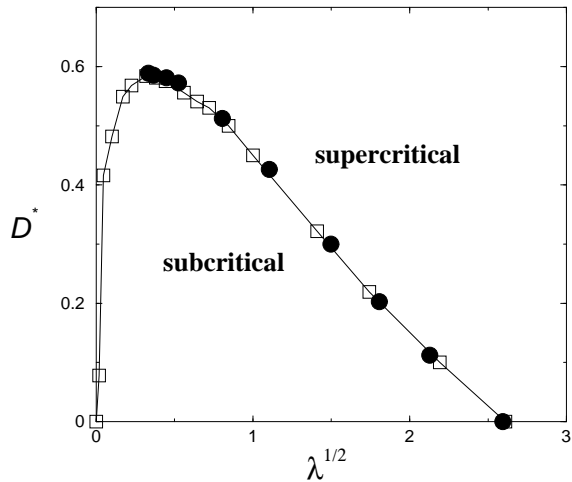}
\caption{Phase diagram for the conserved TAM (full circles) in 
$D^{*}$   versus $\lambda^{1/2}$ space.
The supercritical and subcritical regimes are separated by a critical line..
For comparison, we  show the results obtained by Dickman (squares).}
\label{fig5}
\end{figure}

Using the same  
procedure for nonzero values of $D$ we have built the phase diagram,
shown in Fig \ref{fig5}. We used the Eqs. (\ref{equi}) and (\ref{convert2})
to convert the rates. The  results obtained by Dickman
\cite{dic89a,dick1990}
are also plotted in the same figure for comparison.

Our results are in good agreement with those obtained for the ordinary 
ensemble. For example, the results obtained by Dickman \cite{dick1990} for
the maximum critical diffusion $D^{*}_{\rm max}$ is 0.587 and
its respective critical creation rate  $\lambda_{c}$ is $0.1$.
Our results for this point of the phase diagram 
are $D^{*}_{\rm max}=0.589$ and $\lambda_{c}=0.11$. Concerning
the reentrant phase, the conserved
ensemble seems to be inappropriate to determine the transition
line. In the ordinary ensemble,
isolated particles create new particles leading to
the appearance of triplet even at very small rates and
small number of particles.
In the conserved ensemble, on the other hand, it is necessary the existence of
at least one  triplet for the occurrence of a creation-annihilation
process. 
In the absence of three adjacent particles, only the 
diffusive process will produce a triplet. If the
particles are scattered
the reunion of three particles 
may not occur. In this case the determination of $\alpha$ 
will be affected by  low statistics. 
We remark that 
the reentrant phase diagram for the TAM was confirmed 
by time-dependent numerical simulations \cite{dick1990}, 
which critical exponents belong to the DP universality class.

\begin{figure}
\setlength{\unitlength}{1.0cm}
\includegraphics[scale=0.5]{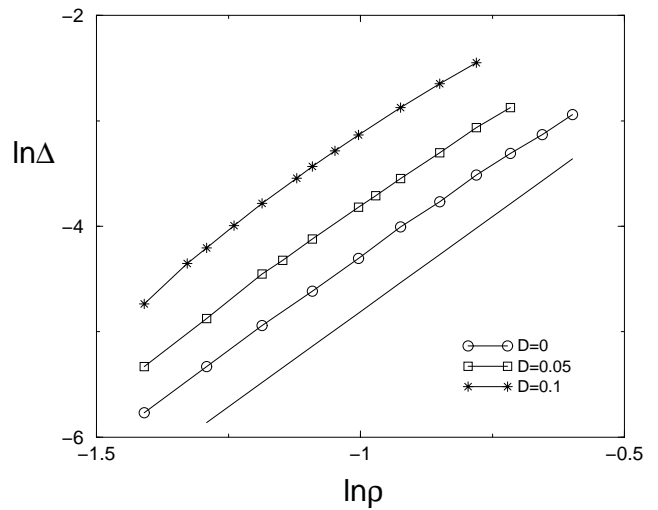}
\caption{log-log plot of $\Delta\equiv\alpha_{c}-\alpha$ versus
the density $ \rho$, in the supercritical regime for the conserved TAM for
some values of diffusion rate. The straight line has a slope 3.61. }
\label{betatam}
\end{figure}

In  Fig. \ref{betatam} we show the log-log plot of 
$\Delta \equiv \alpha_{c}-\alpha$ versus $\rho$  
for some values of diffusion rate in the
supercritical regime using a lattice of size $L=10000$ and 
from $10^{6}$ to $10^{7}$ Monte Carlo steps to evaluate the averages.
For $D=0$, we obtained  $1/\beta=3.60(3)$ 
that is in  agreement with 
the value  $1/\beta=3.61(1)$  obtained for the 
CCP \cite{tome2001}. 
The values for $1/\beta$ obtained from simulations at
$D=0.05$ and $D=0.1$ are also compatible with this value.

\section{Pair contact process}

\subsection{Constant Rate ensemble}

The  ordinary PCP \cite{jensen1993} is a
nonequilibrium model which, like the 
contact process (CP), exhibits a phase
transition to an absorbing state, but differently from  this one,
the PCP possesses infinitely absorbing states. Numerical and theoretical
studies indicate that the PCP (without diffusion) 
also belongs to the DP universality class.
 
The PCP is represented by the chemical reactions:
\begin{equation}
\label{scheme4}
0+A+A \rightarrow  A+A+A,
\end{equation}
\begin{equation}
\label{scheme2}
A+A \rightarrow 0+0,
\end{equation}
describing the  catalytic 
creation of a particle and annihilation of a pair of particles, respectively.
Notice that it is necessary two particles to create a new
particle.

The PCP is defined by the following rules. A pair of neighboring particles is
chosen at random. With probability $p$ it is annihilated and with 
probability $1-p$ a new particle is created in one of its nearest
neighbor sites.  The  dynamics 
 is governed by pairs of neighboring particles, instead of
isolated particles. As a consequence, any configuration absent of
pairs of neighboring particles
is absorbing.  The order parameter  is the density of pairs of
neighboring particles,
instead of the density of particles.

Several studies \cite{jensen1993,jensend1993,dic02,dic1998}
show that the one dimensional PCP exhibits 
a second order transition to an absorbing state at $p_{c}=0.077090(5)$
\cite{dic1998}.
Close to the critical point the density of pairs of neighboring
particles $\rho_{p}$
follows the power law behavior with
an  exponent $\beta=0.2765$ \cite{dic02}. The exponents
$\delta$, $\eta$ and $z$ were also obtained for the PCP
\cite{jensend1993} and they are compatible with the typical values of
the DP universality class. 

Our aim here consists in showing that the PCP 
can be also described by a dynamics that conserves the particle number.
To compare our results with  ordinary versions, we should note that
the strengths
of creation and annihilation rates are $k_c$ and $k_a$
are proportional to $1-p$ and $p$, respectively. Therefore
$\alpha=k_a/k_c=p/(1-p)$ from which follows the
relation between the parameters $p$ and $\alpha$ used here
\begin{equation}
\label{p}
p=\frac{\alpha}{\alpha+1}.
\end{equation}

\subsection{Constant particle number ensemble}

The Monte Carlo simulation  of the conserved PCP 
without diffusion is performed as follows:
A pair of neighboring particles  is selected at random. Next, two
active sites are chosen at random. 
Here,  active sites are empty sites
surrounded by at least a pair of neighboring particles at the same side.
The particles belonging  to the selected pair jump to the two chosen 
active sites.
The rate $\alpha$ is evaluated using Eq. (\ref{equiv}) where the
creation  rate is given by
\begin{equation}
\omega_{i}^{c}=(1-\eta_{i})\frac{1}{z}\sum_{\delta}\eta_{i+\delta}
\eta_{i+2\delta},
 \end{equation}
and the annihilation rate $\omega_i^{a}$
 is given by the expression (\ref{rate2}).

\begin{figure}
\setlength{\unitlength}{1.0cm}
\includegraphics[scale=0.48]{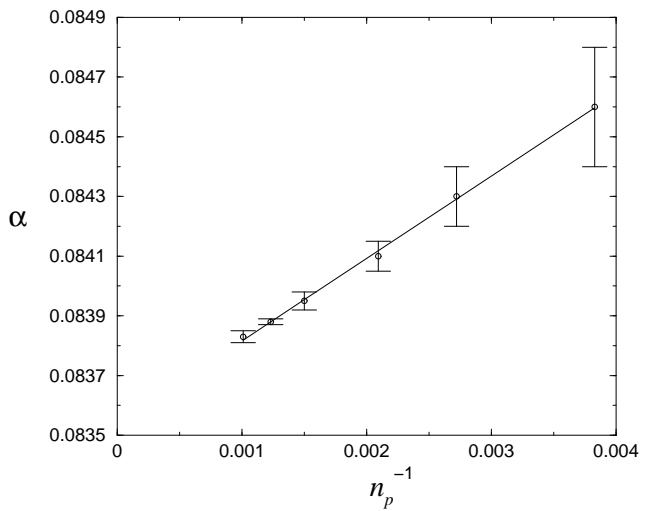}
\caption{Values of $\alpha$ versus the number of pairs of neighboring
particles $n_{p}$  for the conserved PCP
in the absence of diffusion
for an infinite system. The line corresponds to the linear 
extrapolation using Eq. (\ref{expsub}).}
\label{CPCP}
\end{figure}

\begin{figure}
\setlength{\unitlength}{1.0cm}
\includegraphics[scale=0.48]{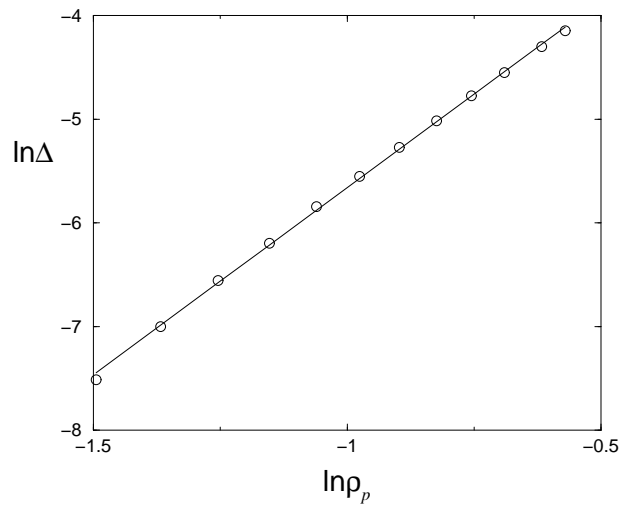}
\caption{Log-log plot of $\Delta\equiv\alpha_{c}-\alpha$ versus
the density of pairs of neighboring particles $\rho_p$,
 in the supercritical regime for the conserved PCP in
the absence of diffusion.  The straight line has a slope 3.61. }
\label{betapair}
\end{figure}

To compare the results coming from distinct ensembles, we have also simulated 
the ordinary pair contact process in one dimension. 
For both  ensembles, we  used  
a lattice of size $L=10000$ and from $10^{6}$ to $10^{7}$ Monte
Carlo steps to evaluate the averages. For example, simulating
the ordinary version for the value of rate
$\alpha=0.01$ we obtained the mean density of particles $\rho=0.96902(1)$
and the mean density of pairs of neighboring particles $\rho_{p}=0.94920(2)$. 
In the conserved 
version for $\rho=0.96900$ we obtain the averages $\rho_{p}=0.94918(2)$
and $\alpha=0.0100(1)$.
Small discrepancies are due the fact that the system size is finite.

In absence  of diffusion, any configuration 
without pairs of neighboring particles 
is absorbing and, therefore the conserved PCP 
also possesses infinitely  absorbing states.

To determine the critical rate
$\alpha_c$ we have  studied the system in the 
subcritical regime. Since a configuration absent of pairs of
neighboring particles is absorbing,
during the simulations  the system may fall into
an absorbing state. Whenever this happens, we  allow isolated 
particles to jump to empty sites surrounded by one particle, in order
to ``create'' pairs of neighboring particles. In Fig. \ref{CPCP}
we show the number of pairs of neighboring particles and its respective 
value of $\alpha$ in the subcritical regime. 
An linear extrapolation in $n_{p}^{-1}$
gives $\alpha_{c}=0.08353(5)$. Using Eq. (\ref{p}) we find
$p_{c}=0.07709(5)$ that is in excellent agreement with the value 
$p_{c}=0.077090(5)$ for the ordinary pair contact process \cite{dic1998}.

To calculate the fractal dimension $d_{F}$ for the conserved PCP, we
adopted the following procedure. We  measured
the maximum distance $R$ between 
two pairs of neighboring particles of a subsystem of fixed size $L$.
The  log-log plot 
of $R$ versus $n_{p}$ is shown in Fig. \ref{fractal}. 
The value of the inverse of the fractal dimension
is $1/d_{F}=1.33(1)$, in very good agreement with the value $1.338(6)$
\cite{sabag}.

Fig. \ref{betapair} shows a log-log plot of 
$\Delta\equiv\alpha_{c}-\alpha$ versus the density $\rho_p$.
The slope of the straight lines fitted to the data points for
the conserved PCP has
slope  3.61(3), in good agreement 
with the value $3.61(1)$, obtained  for the CCP.

\section{Conclusion}

We  have analyzed, by numerical simulations, three one-dimensional 
nonequilibrium models in which particles 
belonging to a  cluster of $\ell$ particles  jump to $\ell$ 
distinct active sites. They are conserved versions of models
originally defined in the constant rate ensemble.
Our approach is general in the sense that any process 
in which the process of creation and annihilation of particles are 
mutually exclusive can be described by a dynamics that conserves the 
number of particles. Our results for the three models studied here
show  that not only 
universal quantities but also nonuniversal parameters  are in excellent 
agreement with their respective ordinary versions.

\section*{ACKNOWLEDGMENT}

C. E. F. thanks the financial support from 
Funda\c c\~ao de Amparo \`a Pesquisa do
Estado de S\~ao Paulo (FAPESP) under Grant No. 03/01073-0.

\end{document}